# Only the Bad Die Young: Restaurant Mortality in the Western US

Tian Luo*
U.S. Bureau of Labor Statistics, San Francisco, CA

Philip B. Stark
Department of Statistics, University of California, Berkeley

DRAFT 30 October 2014

*Abstract*

Do 9 out of 10 restaurants fail in their first year, as commonly claimed? No. Survival analysis of 1.9 million longitudinal microdata for 81,000 full-service restaurants in a 20-year U.S. Bureau of Labor Statistics non-public census of business establishments in the western US shows that only 17 percent of independently owned full-service restaurant startups failed in their first year, compared with 19 percent for all other service-providing startups. The median lifespan of restaurants is about 4.5 years, slightly longer than that of other service businesses (4.25 years). However, the median lifespan of a restaurant startup with 5 or fewer employees is 3.75 years, slightly shorter than that of other service businesses of the same startup size (4.0 years).

# Introduction

A 2003 television advertisement claimed that 9 out of 10 restaurant startups fail in their first year.[1] Sample-based and local studies since then have found much lower failure rates for restaurants in their first year (e.g., Parsa et al. 2005). Yet many still assert that the restaurant industry has some of the highest business failure rates among all types of businesses. The belief that restaurants are particularly risky startup businesses seems pervasive among entrepreneurs and lenders.

At the same time, the number of restaurant establishments has grown steadily by about 2 percent per year over the past decade.[2] Restaurants are a significant part of American life. Through good and bad economic times, Americans have eaten out at about the same rate for several decades. We eat out for many reasons, including convenience, ambience, and fine dining (Ashima & Graubard, 2004; Park, 2004; Warde & Martens, 2000). The average American household spends 5 to 6 percent of its income on food away from home, over $50 a week per household on average.[3]

Using a microdata extract from the U.S. Bureau of Labor Statistics (BLS) Quarterly Census of Employment and Wages (QCEW) program, we examine whether independently owned full-service restaurants are particularly risky businesses. This longitudinal and essentially complete data set allows business survival rates to be calculated with low bias and no sampling error.

This study uses a *longitudinal census* of businesses. Typical studies are sample-based and are either local or have relatively small sample sizes, which makes extrapolation to the US highly uncertain, even regionally. Longitudinal studies of business mortality are typically cohort-based. Restricting analysis to a cohort controls for some sources of confounding, but limits sample sizes and exacerbates other sources of confounding. For instance, exogenous effects such as macroeconomic shocks confound with the underlying survival function. In contrast, we estimate survival functions for all businesses born in a 20-year period.

We study single-establishment and independently owned full-service restaurants as well as all other single-establishment service-providing businesses in the western US. Multi-establishment and "chain" restaurants are excluded because their operational structure is different, which may affect their survival. Below, we refer to single-establishment full-service restaurants as "restaurants." We use the terms "restaurant establishments," "restaurant businesses," and "restaurants" interchangeably: for independently owned, single-establishment restaurants, there is no distinction. Our analysis compares restaurants to other single-establishment start-up businesses.

---

[1] NBC broadcast a program titled "Restaurant: A Reality Show" with an American Express commercial in 2003.
[2] U.S. Bureau of Labor Statistics, Quarterly Census of Employment and Wages, NAICS 722110, http://www.bls.gov/cew
[3] U.S. Bureau of Labor Statistics, Consumer Expenditure Survey, "Table 52. Region of residence: Shares of average annual expenditures and sources of income," years 1990 through 2011, http://www.bls.gov/cex/csxshare.htm

This paper has two goals: 1) Illustrate nonparametric methods for estimating and comparing business survival functions and 2) compare survival rates of restaurants and other businesses to test the common belief that restaurants have higher failure rates.

## Data

This paper analyses an extract of the BLS Quarterly Census of Employment and Wages (QCEW) longitudinal database from 1992 to 2011[4] for eight western states in the US.[5] These monthly data are compiled quarterly for state unemployment insurance tax purposes, edited, and submitted to the BLS. QCEW is a cooperative program among BLS and the State Workforce Agencies. The program collects information reported by employers covering approximately 98 percent of jobs in the United States.[6] Coverage for single-establishment full-service restaurants is essentially complete, to the extent that those businesses comply with federal reporting requirements.

The extract contains 136 million observations of over 5.8 million establishments in both private and public sectors. Each observation represents one business establishment in one quarter. Public sector establishments, private households[7], and multi-establishment businesses[8] are excluded from our analysis. Comparisons between restaurants and other businesses include only on service-providing businesses, excluding utilities. Of the 1.9 million single-establishment service-providing businesses in this dataset, 81,000 were full-service restaurants. Employment in these single-establishment restaurants comprised 68.5 percent of the employment by all full-service restaurant (including multi-establishment or chained restaurants) and 2.3 percent of all private sector employment in the western US.

We take an establishment's birthdate to be the first quarter in which it appears in this database after Q1 1992 and had positive employment. We consider an establishment to have died in the last quarter in which it had positive employment before disappearing from the database. These definitions of birth and death dates are consistent with previous studies (Sadeghi, 2008; Spletzer, 2000). An address change is not considered a death: businesses are tracked, even if they move. Mergers and spinoffs are recorded and treated as censoring in the analysis.

---

[4] The original data extract includes 1990, 1991, and 2012. We excluded those years because reported births and deaths were unreliable as the database was brought online in the earlier years and because revisions to the fields used to determine birth and deaths ("first positive employment date" and "last positive employment date") introduced a year lag from the most recently observed data.
[5] Western US States include Alaska, Arizona, California, Hawaii, Idaho, Nevada, Oregon, and Washington.
[6] Employment and "establishments" not in this dataset include self-employed workers, most agricultural workers on small farms, all members of the Armed Forces, elected officials in most states, most employees of railroads, some domestic workers, most student workers at schools, and employees of certain small nonprofit organizations. (http://www.bls.gov/cew/cewfaq.htm) Furthermore, establishments are excluded if they are not covered under unemployment insurance (UI) which depends on state laws. For example, in California, any establishment paying $100 or more in wages in a quarter is required by law to register for UI. (http://www.edd.ca.gov/payroll_taxes/am_i_required_to_register_as_an_employer.htm) Thus, the extent to which undercoverage occurs (when an establishment pays out less than $100 per quarter, or when it does not comply with federal reporting regulations), is expected to be small.
[7] Private households, NAICS 814110, includes maids, nannies, cooks, butlers, and gardeners
[8] Multi-establishment businesses have higher survival rates but are excluded from the analyses in this paper.

We use data on all establishments born between 1992 Q2 and 2011 Q2. Establishments fall in four groups, illustrated in Figure 1: 1) birth and death dates known, 2) birth date known but death unknown (right-censored), 3) birth date unknown and death date known, and 4) both birth and death dates unknown. Our estimates use type 1 and 2 observations. We know of no unbiased method to incorporate observations of types 3 and 4, even though those observations still give a lower bound on the age at death. Their presence in the data is conditional on survival beyond 1992, but their ages in 1992 are not known. Those data comprise right-censored samples from different, unknown conditional distributions of lifetimes.

**Figure 1.** Types of observations

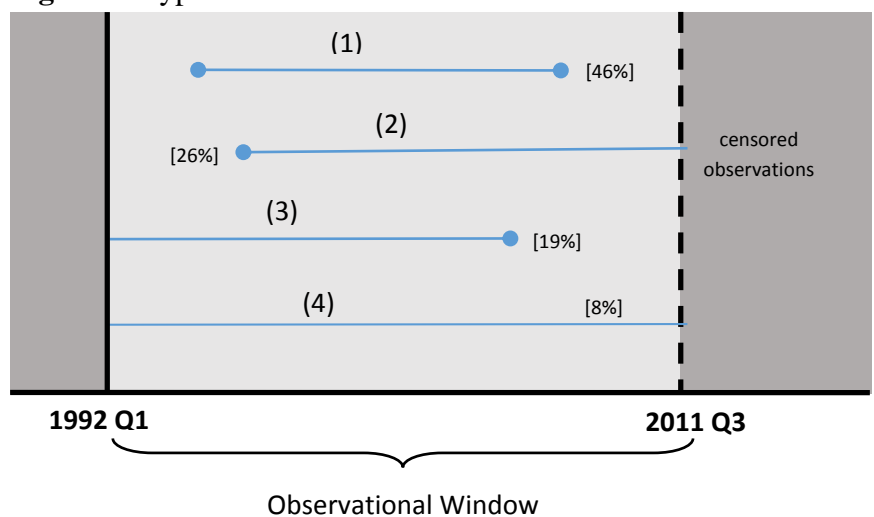

Note: ovals at ends of lines indicate observed births and deaths. Observation type is in parenthesis. Percentage of all observations is in square brackets. Type 2 observations are right-censored. Our estimates use type 1 and 2 observations.

**Methodology**

Since the data are a census, sampling bias, to the extent that it occurs, results from undercoverage—i.e., establishments missing from the census. Sampling bias is expected to be negligible, especially for the establishments discussed in this paper.[9] But for undercoverage, standard (albeit unverifiable and somewhat contrived) stochastic assumptions suffice to ensure that the survival function can be estimated with low bias:

> 1) Survival times of different establishments are random, independent, and identically distributed. In particular, the distribution does not depend on whether the establishment is born early or late in the observational window.

---

[9] As mentioned above, coverage for these establishments is complete but for restaurants that illegally fail to report.

2) Censoring is independent of survival time.

The assumption that failures are random at all is an epistemic leap. The assumption that failures are independent ignores correlations that might result from local economic effects at the scale of neighborhoods and up. Under assumptions (1) and (2), the nonparametric maximum likelihood estimator of the survival function is the Kaplan-Meier (KM) Product Limit estimator (Kaplan and Meier, 1958):

$$\hat{S}(t) = \prod_{t_i \leq t} (1 - \frac{d_i}{n_i}),$$

where $\hat{S}(t)$ is the estimated probability of surviving past time $t$, $n_i$ is the number of establishments that are "at risk" at time $t_i$, and $d_i$ is the number of deaths at time $t_i$. The number at risk at time $t_i$ is

$$n_i = s_i - c_i,$$

where $s_i$ is the number of firms still alive just prior to time $t_i$, and $c_i$ is the number of cases censored after $t_{i-1}$ but before $t_i$.

The KM estimator uses establishments that are born and die within the observational window, as well as right-censored observations (observations of type 2). Under assumptions (1) and (2), the KM estimator is consistent and asymptotically unbiased.

*Censored observations.* The observations end in Q3 2011, so survival times of establishments still alive then are censored: we do not know how much longer they live. If an establishment changed ownership or merged with or was acquired by another business, we treat it as (right) censored at the date of this change. The KM estimator deals with censoring by eliminating censored cases from the "at risk" group rather than treating them as deaths.

Under the stochastic assumptions, the estimated variance of $\hat{S}(t)$ is

$$\widehat{Var}[\hat{S}(t)] = [\hat{S}(t)]^2 \sum_{t_i \leq t} \frac{d_i}{n_i(n_i - d_i)}.$$

Estimated variances of survival rates reported in this paper are very small because the number of observations is large. We omit variances and confidence intervals for survival rates, although we do consider uncertainty when we test the hypothesis that two groups share the same survival function.

To compare two survival functions, we use a logrank test or Mantel–Cox test (Nathan, 1996; Peto, 1972; Harrington, 2005) to test the hypothesis that two underlying survival functions are the same. This test compares the expected and observed number of deaths between two groups at each observed event time. The logrank statistic is

$$Z = \frac{\sum_{i=1}^{T}(d_{1i} - n_{1i}\frac{d_i}{n_i})}{\sqrt{\sum_{i=1}^{T}\frac{n_{1i}n_{2i}d_i(n_i - d_i)}{n_i^2(n_i - 1)}}},$$

where $d_j$ and $n_i$ are the number of deaths and number at risk, respectively, for both groups at time $i$, and $d_{1i}$ and $n_{1i}$ are the number of deaths and number at risk, respectively for group 1 at time $i$. Under the null hypothesis that the two groups are independent and have the same hazard function, the distribution of the logrank statistic is approximately standard normal; we confirmed that the normal approximation was accurate by simulation (simulations using 10,000 replications typically agreed with the normal approximation to within 0.001, which is on the order of the sampling error).

## Birthrates over time

As shown in Figure 2, other than seasonality, there does not appear to be a pattern to birthrates and deathrates between 1992 and 2011, so the census contains roughly equal numbers of businesses born in each year. However, different cohorts affect different parts of the survival curve, since businesses born late in the window do not contribute to the estimate of long-term survival rate. (We understand that the downward and upward spikes in birthrates in 1997 and 1998 are likely caused by administrative changes that affect the reporting requirements of businesses covered under Unemployment Insurance.)

**Figure 2.** Quarterly birth and death rates of service-providing businesses and restaurants

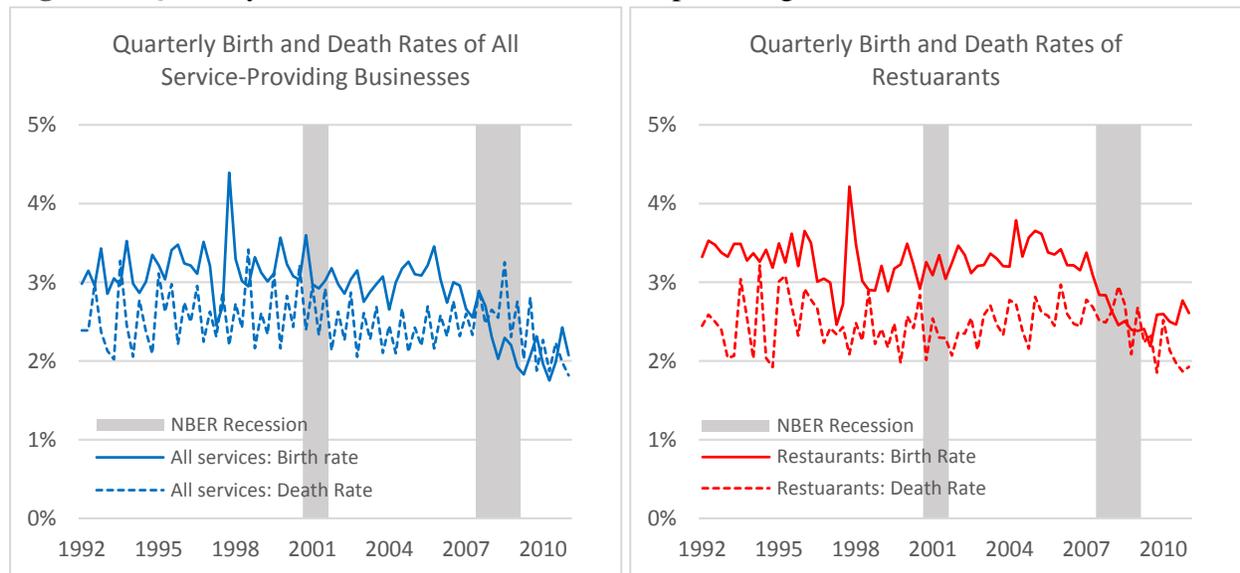

## Startup size and time of birth

We first examine differences in survival rates among startups of different sizes.[10] Survival rates tend to increase with startup size (Figure 3). Compared to businesses that started with 5 or fewer employees (small), startups with 6 to 20 employees (medium) had an annual survival rate 1.6 percent higher, while those with 21 or more (large) had an annual survival rate 2.8 percent higher. Twenty-one percent of small startups survived past age 15, in contrast to 27 percent of medium startups and 33 percent of large startups. Larger startups may need more capital, but tend to have higher survival rates.

**Figure 3.** Survival rates of service-providing businesses born after Q1 1992, grouped by number of employees at birth

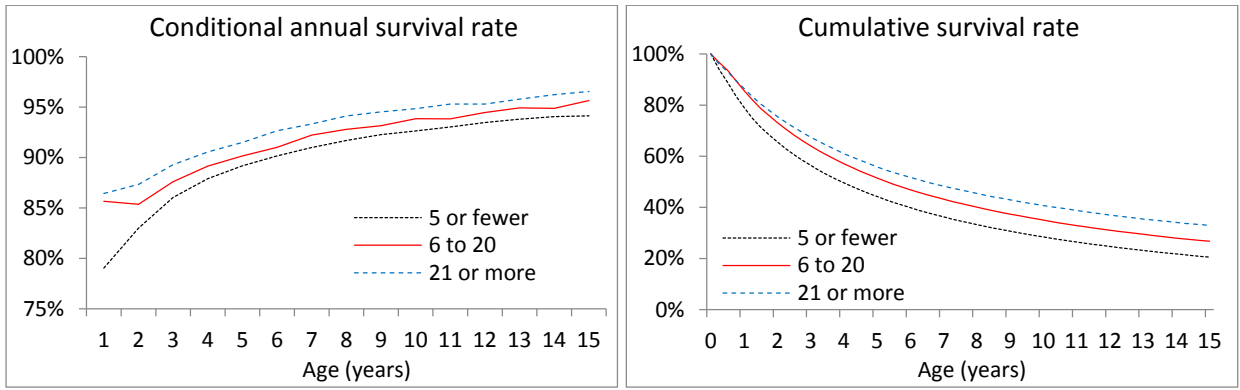

To check whether the stationarity assumption is obviously violated, we examine whether date of birth is related to longevity. Figure 4 shows only negligible differences among survival rates of establishments born in different phases of economic cycles. Over the course of 10 years, survival rates for establishments born in different phases of economic cycles have survival rates that differ by less than 2 percentage points.

---

[10] Startup size is defined as the number of employees at birth (the first month of positive employment)

**Figure 4.** Survival rates of service-providing businesses by birth year

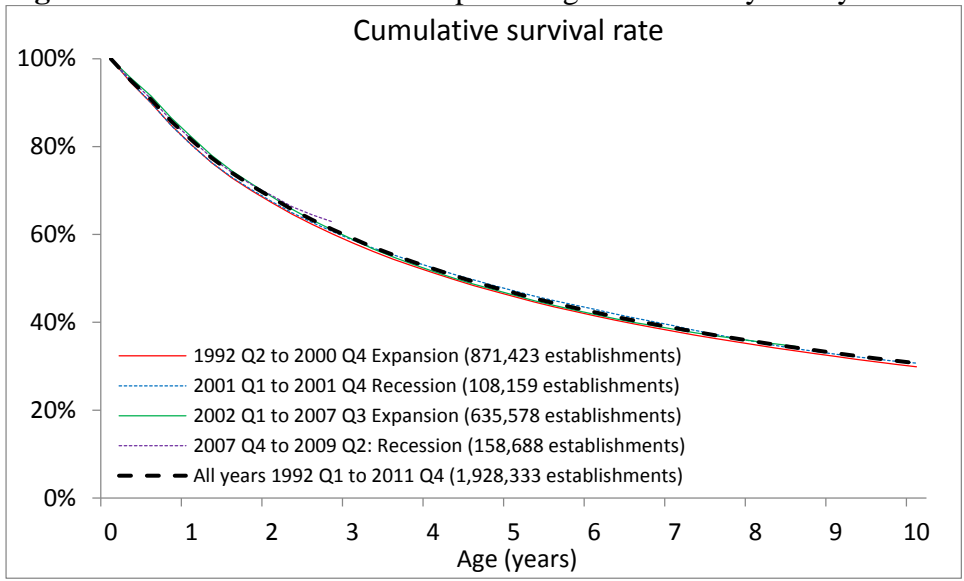

## Restaurants versus other service-providing businesses

In stark contrast to the commonly cited statistic that 90 percent of restaurants fail in their first year, only about 17 percent of restaurants failed in the first year—*lower* than the average first-year failure rate of 19 percent for all other service-providing businesses. In the Western US, restaurants and other service-providing businesses have median lifetimes of roughly 4.5 and 4.25 years, respectively. Figure 5 shows the survival rate of restaurants and service-providing businesses as well as the conditional quarterly survival rate. Restaurant startups tend to have slightly higher survival rates than other service startups. The difference in survival functions between restaurants and all other services is highly statistically significant (table 2).

The quarterly conditional survival rates (probability of surviving a given quarter given the establishment was alive at the beginning of that quarter) are fairly high in the first year, drop during the second year, and then increase in a concave fashion. Previous studies have also found that survival rates generally increase with age (Evans, 1987; Popkin, 2001). Furthermore, the *liability of adolescence* argument also suggests that the survival rate for a firm's initial year is higher because businesses generally can survive for a year on initial resources (Brüderl et al., 1992). As a result, conditional survival rates as a function of age tend to be U-shaped in early years (Bruderl and Schussler, 1990; Fichman and Levinthal, 1991).

**Figure 5.** Survival rates of service-providing businesses and restaurants

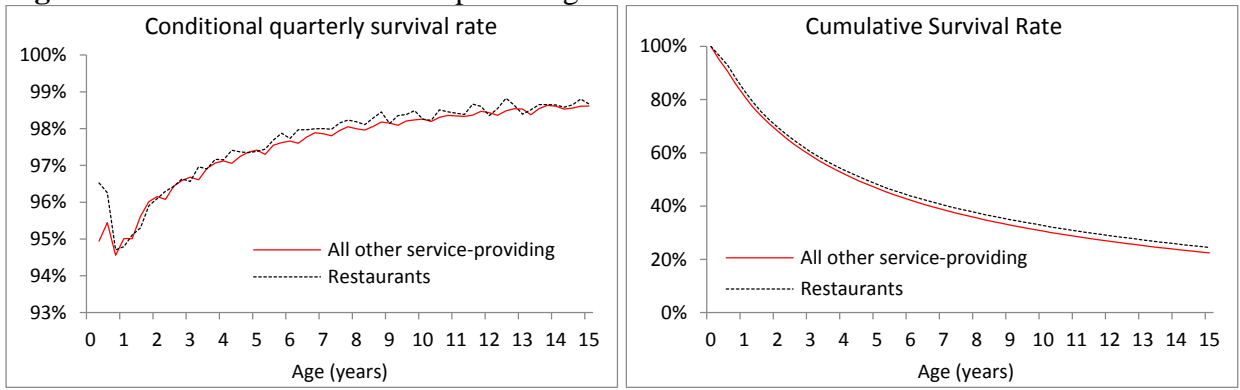

As shown in the previous section, larger startups generally have higher survival rates. Date of birth may produce some variation as well. Table 1 compares survival rates between restaurants and other businesses for various startup sizes and birth epochs. In each size group, the difference in survival rates between restaurants and other service establishments is about the same, regardless of birth period. Hence, the assumption of stationarity seems unlikely to bias the KM estimates.

In every birth period, restaurants as a group have predominantly slightly higher survival rates than other service businesses. However, risk does depend on size at birth. The median lifespan of restaurants that started with 20 or fewer employees is about 3 months *shorter* than other businesses of the same startup size, but restaurants with 21 or more employees had median lifespan about 9 months *longer* than other businesses with the same startup size.

**Table 1.** Difference in median lifetime (years) between restaurant and all service-providing businesses, by startup size and birth period

| Employment at birth | 1992-2000 Expansion | 2001 Recession | 2002-2007 Expansion | All birth years, 1992-2011 |
|---|---|---|---|---|
| 5 or fewer | **-0.25** | 0.25 | -0.25 | **-0.25** |
| 6 to 20 | 0.00 | -0.50 | **-0.75** | **-0.25** |
| 21 or more | **2.00** | 0.25 | **1.75** | **0.75** |
| All sizes | **0.50** | **0.25** | 0.00 | **0.25** |

Note: bold numbers indicate survival rates significantly different between restaurants and other businesses at the 5% significance level for the complete survival function.

## Restaurants versus other startups

Of over 500 different types of single-establishment service startups (by 6 digit NAICS) between 1992 and 2011, full-service restaurants rank the highest[11] in number of startups in the western US. The top 15 startup categories made up about one-third of all startups (see table 2).

---

[11] Wholesale Trade Agents and Brokers (NAICS 425120) are excluded from this count as this NAICS industry coding was established in 2007, combining 68 different 6-digit previous (2002) NAICS code industries consisting of wholesalers in a wide range of industries.

Some of the most popular startups with the worst survival rates include janitorial services and custom computer programming services, which have median lifetimes below 3.5 years. Popular businesses with best survival rates provide professional services: offices of dentists, physicians, and lawyers, which have median lifetime of over 19.5 years, 10.75 years, and 7.5 years, respectively.

**Table 2.** Survival rates by top startups

|   | Industry | NAICS | 1-year survival rate | Median lifetime | Median lifetime Small only[1] | \|z-stat\| | p-value[2] | Number of observed businesses |
|---|---|---|---|---|---|---|---|---|
|   | **All service-providing (excluding restaurants)** | **42 to 81** | **0.81** | **4.25** | **4.00** | **12.98** | **<.0001** | **1,846,900** |
| 1 | Full-service restaurants | 722110[3] | 0.83 | 4.50 | 3.75 | - | - | 81,500 |
| 2 | Limited-Service Restaurants | 722211[4] | 0.81 | 3.75 | 3.00 | 20.00 | <.0001 | 51,400 |
| 3 | Offices of Real Estate Agents and Brokers | 531210 | 0.79 | 3.50 | 3.50 | 30.17 | <.0001 | 46,000 |
| 4 | Offices of Physicians | 621111 | 0.90 | 10.75 | 10.00 | 64.42 | <.0001 | 46,000 |
| 5 | Offices of Lawyers | 541110 | 0.87 | 7.50 | 7.00 | 37.44 | <.0001 | 42,600 |
| 6 | Other Scientific and Technical Consulting Services | 541690 | 0.81 | 3.50 | 3.25 | 20.37 | <.0001 | 38,900 |
| 7 | Custom Computer Programming Services | 541511 | 0.80 | 3.25 | 3.00 | 31.87 | <.0001 | 38,000 |
| 8 | Computer Systems Design Services | 541512 | 0.78 | 3.25 | 3.00 | 34.50 | <.0001 | 32,200 |
| 9 | Insurance Agencies and Brokerages | 524210 | 0.83 | 6.00 | 5.50 | 17.58 | <.0001 | 30,800 |
| 10 | Landscaping Services | 561730 | 0.81 | 4.75 | 4.25 | 0.43 | 0.667 | 27,000 |
| 11 | Administrative Management and General Management Consulting Services | 541611 | 0.78 | 3.50 | 3.00 | 25.88 | <.0001 | 26,900 |
| 12 | Janitorial Services | 561720 | 0.76 | 3.00 | 2.50 | 32.30 | <.0001 | 25,200 |
| 13 | General Automotive Repair | 811111 | 0.81 | 4.50 | 4.25 | 3.43 | 0.001 | 25,100 |
| 14 | Offices of Dentists | 621210 | 0.93 | >19.50[5] | >19.50 | 86.31 | <.0001 | 23,100 |
| 15 | Engineering Services | 541330 | 0.83 | 5.25 | 4.50 | 11.52 | <.0001 | 22,300 |

[1] Startup size of 5 or fewer
[2] *P*-value of the hypothesis that the survival curve of the subset of businesses in the row is identical to the survival curve of restaurants of all startup sizes.
[3] NAICS changed to 722511 in 2012
[4] NAICS changed to 722513 in 2012
[5] Due to a high rate of survival and large number of right-censored cases, median lifetime was not reached by the end of the observational period of 20 years

Table 3 compares restaurant startups with other selected startup businesses in industries such as retail trade and services, including professional, administrative, educational, amusement, repair, and personal. Restaurant survival rates are roughly in the middle. Businesses with low first-year survival and median lifetimes include record stores, computer training, amusement arcades, and photofinishing stores. These businesses had a median lifetime of 3 years or less and a first-year survival of less than 4 in 5. At the other end, musical instrument stores, pet care services, and convenience stores had some of the best survival rates, with median lifetime of more than 5 years. Businesses with survival rates near those of restaurants include bars, building material supply stores, automobile driving schools, sewing stores, and drycleaning and laundry services.

Table 3 also shows survival rates for other startups of similar sizes. The relative survival rates (compared to restaurants) are similar for startups of similar sizes. Table 3 also shows the median lifetime of small startups. For example, automotive repair shops and hair and skin-care services had similar survival rates as restaurant startups of similar size. However, across all startup sizes, these businesses had significantly lower median lifetime than that of restaurants.

**Table 3.** Survival rates by selected industries

| Industry | NAICS | 1-year survival rate | Median lifetime | Median lifetime Small only[1] | \|z-stat\| | p-value[2] | Number of observed businesses |
|---|---|---|---|---|---|---|---|
| Record stores | 451220 | 0.77 | 2.50 | 2.25 | 22.34 | <.0001 | 1,500 |
| Computer training | 611420 | 0.77 | 3.00 | 2.75 | 13.34 | <.0001 | 1,400 |
| Amusement arcades | 713120 | 0.80 | 3.00 | 2.25 | 5.16 | <.0001 | 500 |
| Photofinishing | 81292 | 0.79 | 3.00 | 2.75 | 14.28 | <.0001 | 1,100 |
| Hobby, toy, and game stores | 45112 | 0.81 | 3.25 | 3.25 | 13.01 | <.0001 | 3,200 |
| Electronics and appliance stores | 443 | 0.81 | 3.50 | 3.25 | 23.52 | <.0001 | 18,100 |
| Packaging and labeling services | 561910 | 0.81 | 3.50 | 2.75 | 5.48 | <.0001 | 1,000 |
| Household goods repair and maintenance | 8114 | 0.78 | 3.50 | 3.25 | 12.68 | <.0001 | 7,000 |
| Clothing and clothing accessories stores | 448 | 0.83 | 3.75 | 3.50 | 12.92 | <.0001 | 17,800 |
| Book stores | 451211 | 0.82 | 3.75 | 3.25 | 8.06 | <.0001 | 2,700 |
| Electronic equipment repair and maintenance | 81121 | 0.80 | 3.75 | 3.25 | 12.19 | <.0001 | 5,600 |
| Specialty food stores | 4452 | 0.82 | 4.00 | 3.50 | 6.19 | <.0001 | 10,500 |
| Cosmetic and beauty supply stores | 451 | 0.83 | 4.00 | 3.50 | 15.40 | <.0001 | 28,400 |
| Photography studios | 541921 | 0.81 | 4.00 | 4.00 | 6.99 | <.0001 | 3,300 |
| Locksmiths | 561622 | 0.79 | 4.00 | 3.50 | 1.88 | 0.060 | 1,100 |
| Furniture and home furnishings stores | 442 | 0.83 | 4.25 | 4.00 | 9.68 | <.0001 | 15,300 |
| Sporting goods stores | 45111 | 0.85 | 4.25 | 4.00 | 4.29 | <.0001 | 7,500 |
| Automotive repair and maintenance | 8111 | 0.81 | 4.25 | 3.75 | 10.79 | <.0001 | 50,400 |
| Hair, nail, and skin care services | 81211 | 0.82 | 4.25 | 3.75 | 12.60 | <.0001 | 28,200 |
| Drinking places, alcoholic beverages | 722410 | 0.83 | 4.50 | 4.25 | 3.47 | 0.001 | 10,000 |
| Building material and garden supply stores | 444 | 0.84 | 4.75 | 4.25 | 2.38 | 0.017 | 12,900 |
| Automobile driving schools | 611692 | 0.83 | 4.75 | 4.50 | 0.46 | 0.647 | 900 |
| Sewing, needlework, and piece goods stores | 45113 | 0.87 | 5.00 | 4.50 | 0.63 | 0.528 | 1,600 |
| Drycleaning and laundry services | 8123 | 0.85 | 5.00 | 4.50 | 1.12 | 0.263 | 10,400 |
| Convenience stores | 44512 | 0.84 | 5.75 | 5.00 | 7.50 | <.0001 | 7,500 |
| Pet care services (except veterinary) | 812910 | 0.84 | 5.75 | 5.00 | 5.94 | <.0001 | 3,800 |
| Musical instrument and supplies stores | 45114 | 0.88 | 6.00 | 5.75 | 2.97 | 0.003 | 1,000 |

[1] Startup size of 5 or fewer
[2] Test of equality of survival compared with restaurants (all startup sizes)

## Discussion

Many factors affect the survival of a restaurant business, including organizational factors and strategies (Mahmood, 1991; Agarwal and Audretsch, 2001; Audretsch, 1994); finances, marketing, and product mix; and environmental conditions such as resource availability and competition (Romanelli, 1989). Marketing is important, as are location, food quality, and the characteristics of owner or manager. Successful restaurants not only have well-defined food products but also an operating philosophy that encompasses business operations and employee and customer relations (Parsa et al., 2005). Restaurants with low earnings and high liabilities are more likely to go out of business (Gu, 2002). Some small restaurants fail in part due to family demands such as divorce, health problems, and retirement; there is evidence that owners of successful restaurants are good at balancing personal and work lives or were not married (Parsa et al., 2005). Survival rates may also vary geographically due to differences in spending habits, taxation, and regulation (O'Neill and Duker, 1986; Edmunds, 1979). This paper restricts attention to survival rates in eight western US states, aggregated across single-establishment businesses of all sizes in all settings—urban, suburban, and rural.

Perhaps due to the visibility and volume of restaurant startups, the public perception is that restaurants often fail. However, as shown in this paper, restaurant turnover rates are not very different from startups of many other different industries.

## Conclusion

The first-year failure rate of single-establishment restaurants in the Western US in the past two decades was about 17 percent, belying the urban myth that 90 percent of restaurants fail in their first year. This first-year failure rate was significantly *lower* than the 19 percent rate of all other service-providing businesses. The median lifetime for restaurants is 4.5 years, slightly more than the 4.25 years for all other service-providing businesses. Many types of businesses have far lower survival rates than restaurants have. Offices of professionals such as dentists, physicians, and lawyers have far higher survival rates than other types of startups.